\documentclass{article}
\usepackage{amsmath,amssymb}
\usepackage{graphicx}
\usepackage[english]{babel}
\usepackage{cite,citesort}

\newcommand{\gref}[1]{(\ref{#1})}
\newcommand{\td}{\text{d}}
\newcommand{\bra}[1]{\left\langle #1 \right|}
\newcommand{\ket}[1]{\left| #1 \right\rangle}
\newcommand{\bk}[2]{\left\langle #1 | #2 \right\rangle}

\newcommand{\de}[2]{\frac{\text{d} #1}{\text{d}#2}}
\newcommand{\dde}[2]{\frac{\text{d}^2 #1}{\text{d} #2 ^2}}
\newcommand{\be}{\begin{equation}}
\newcommand{\ee}{\end{equation}}
\newcommand{\p}[2]{\frac{\partial #1}{\partial #2}}
\newcommand{\pp}[2]{\frac{\partial^{2} #1}{\partial #2^{2}}}
\newcommand{\ppp}[3]{\frac{\partial^{2} #1}{\partial #2 \partial #3}}

\title{Automatic Generation of Matrix Element Derivatives for Tight Binding Models}
\author{Alin M. Elena\footnote{E-mail: Alin.Elena@qub.ac.uk} and Matthias Meister\\
Atomistic Simulation Centre,\\ Department of Physics and Astronomy,\\
  Queen's University Belfast \\ Belfast BT7 1NN, UK, EU} 
\date{\today}
\begin{document}
\maketitle

\begin{abstract}
Tight binding (TB)  models are one approach to the quantum mechanical many particle problem. 
An important role in TB models is played by hopping and overlap matrix elements between the orbitals
on two atoms, which of course depend on the relative positions of the atoms involved. This dependence 
can be expressed with the help of Slater-Koster parameters, which are usually taken from tables. 
Recently, a way to generate these tables automatically was published.  
If TB approaches are applied 
to simulations of the dynamics of a system, also derivatives of matrix elements can appear. In this
work we give general expressions for first and second derivatives of such matrix elements. 
Implemented in a computer program they obviate the need to type all the required derivatives of 
all occuring matrix elements by hand. 
\end{abstract}

\section{Introduction}
Tight binding is an approach to the quantum mechanical many particle problem particularly valuable in cases 
where exact analytic solutions are not available (i.e. almost always) and other numerical approaches 
like CI or DFT are too time-consuming. For an overview see e.g. \cite{FINNIS,Goringe1997}. 
The general structure of the TB 
equations can be derived from DFT \cite{FINNIS,Foulkes1989,Frauen2000}. 
As is not uncommon in quantum mechanics, matrix elements of 
operators and overlap matrix elements between states occur also in TB. 
The distinctive feature of TB is that these matrix elements are considered to arise from atomic-{\it like}
orbitals. The precise way in which these matrix elements are obtained defines the type of TB approach.
The matrix elements can be actually calculated from orbitals localised at atoms, with the orbitals obtained 
in some way before. Another possibility is to consider the matrix elements as disposable parameters 
(empirical TB). The matrix elements depend on the relative positions of the atoms involved. Regardless whether
the matrix elements are obtained from atom-centred orbitals or introduced as parameters, it is important that 
this dependence on the relative position has the basic characteristics brought about by the properties of
atomic orbitals. In particular, the angular dependence of the overlap of two orbitals located at different 
atoms is determined by the angular momentum quantum numbers of the orbitals. For two-centre matrix elements, 
i.e. those which depend on the positions of two atoms only, this angular dependence can be expressed in terms
of Slater-Koster coefficients \cite{Slater1954}, which are published in various 
tables, e.g. \cite{Slater1954,Sharma1979}. Many TB schemes
actually restrict to two-centre integrals, neglecting matrix elements with three or more centres. For a 
discussion of this see for example \cite{FINNIS}.
With increasing angular momentum quantum number $l$ the length of the corresponding Slater-Koster tables and
also the length of individual entries in such tables increase rapidly. Therefore a procedure to calculate 
the Slater-Koster coefficients automatically is very useful. Such an approach has been presented in 
\cite{Podolskiy2004}.
\newline
If the TB approach is applied to studies of the dynamics of a system (e.g. a molecule), also derivatives of the
matrix elements with respect to the positions of the atoms occur, more precisely derivatives of first and 
second order \cite{Todorov2001}. 
To an even higher degree than for the matrix elements themselves, the expressions to
be handled (typed into a computer) quickly become numerous and very complicated with increasing $l$. 
Therefore general expressions for these derivatives are most helpful. In this work, building on 
\cite{Podolskiy2004}, we 
present such expressions, which, while they may still look a bit awkward, can be well implemented in a 
TB code.\newline
In section \ref{ME} we summarise results of \cite{Podolskiy2004} 
and establish the starting point for the subsequent
sections \ref{Fder} and \ref{Sder}, which contain our results for the first and second derivatives,
respectively, of a general two-centre atomic-like matrix element. Some problems with the coordinates
chosen are discussed in section \ref{poles}.   
\section{Matrix elements}
\label{ME}
\begin{figure}[t]
\begin{center}
\includegraphics[width=7cm]{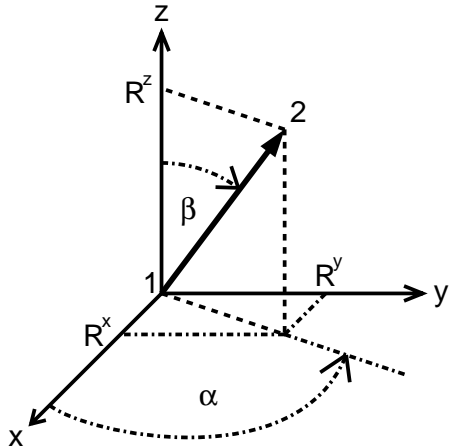}
\end{center}
\caption{The geometric situation. Atom $1$ is located at the origin.
The quantities $R^{x},R^{y},R^{z}$ are the Cartesian coordinates of atom $2$.}
\label{geometry}
\end{figure}

For the convenience of the reader and also to introduce our notation this section summarises results 
of \cite{Podolskiy2004} on the general form of Slater-Koster matrix elements. \\
Let us consider two atoms, 1 at position $\vec{r}_{1}$, 2 at position $\vec{r}_{2}$, and a quantum 
state on each of the atoms, characterised by angular and magnetic quantum numbers \(l_1,m_1\) 
and \(l_2,m_2\), respectively. These quantum numbers determine the 
symmetry properties of the states relevant to the subsequent discussion, whereas other quantum numbers play
no explicit role.
Let $\vec{r}_{12}:=\vec{r}_{2}-\vec{r}_{1}=:|\vec{r}_{12}|\vec{\xi}=:R\vec{\xi}$ be 
the connecting vector of atom 1 and 2. In the two-centre approximation, all matrix elements appearing in 
a tight binding approach depend only on the positions of two atoms and can be rephrased to depend only 
on the connecting vector. Therefore one of the atoms can be taken to be situated in the origin of 
the coordinate system, as illustrated in figure \ref{geometry}. The Cartesian coordinates of 
atom 2 then are
\(R^x,R^y,R^z\), and $\alpha,\beta$ are the Euler angles of the rotation bringing the $z$-axis into 
alignment with the connecting vector. $\alpha$ is defined as the
rotation angle about the z-axis starting from the positive x-axis, $0\le\alpha<2\pi$; $\beta$ gives the
rotation angle about the new y$^{\prime}$-axis (obtained as result of the $\alpha$-rotation), 
starting from the positive z-axis, $0\le\beta\le\pi$. Note that our convention for the first Euler angle 
differs from \cite{Podolskiy2004}. A possible rotation about the connecting vector by a third Euler angle is
irrelevant for our purposes and will not be considered. 

Between the Euler angles and the Cartesian coordinates the following relations hold
(with $R=\sqrt{(R^x)^2+(R^y)^2+(R^z)^2}$ ):
\be
\cos \beta=\frac{R^z}{R},\qquad 
\sin \beta=\sqrt{1-\frac{(R^z)^2}{R^2}}   
\ee
\be
\cos \alpha=\frac{R^x}{\sqrt{(R^x)^2+(R^y)^2}},\qquad 
\sin \alpha=\frac{R^y}{\sqrt{(R^x)^2+(R^y)^2}},\qquad 
if\quad (R^x)^2+(R^y)^2\neq0  
\ee
The case \((R^x)^2+(R^y)^2=0\) implies $\beta=0$ or $\beta=\pi$; $\alpha$ is undefined in this case and 
we will look at this problem in section \ref{poles}. 

The central point is to express a spherical harmonic \(Y_{lm}(\theta,\varphi)\) of
the unrotated coordinates (i.e. $\theta$ is measured from the positive z-axis,
 $\varphi$ from the positive x-axis) as a linear combination of spherical harmonics 
$Y^{(\xi)}_{lm}(\theta^{\prime},\varphi^{\prime})$ in the rotated frame
($\theta^{\prime}$ from positive $\vec{\xi}$-axis, $\varphi^{\prime}$ from
the new x-axis ):
\be
\label{RotCoor}
f(\theta^{\prime},\varphi^{\prime}):= Y_{lm}(\theta(\theta^{\prime},\varphi^{\prime}),\varphi(\theta^{\prime},\varphi^{\prime}))
=\sum_{m^{\prime}}C_{lm}^{m^{\prime}}Y^{(\xi)}_{lm^{\prime}}(\theta^{\prime},\varphi^{\prime})
\ee
Only functions of the same $l$ are involved, because the operator effecting
the rotation by the Euler angles
\be
R(\alpha,\beta,\gamma)
=e^{\frac{i}{\hbar}J_{z^{\prime}}\gamma}e^{\frac{i}{\hbar}J_{\overline y}\beta}e^{\frac{i}{\hbar}J_{z}\alpha}
\ee
(where
$\overline y$ denotes the intermediate y-direction after the $\alpha$-rotation,
$z^{\prime}=\xi$ the final z-direction) commutes with $L^2$ \cite{Daw}.
From \gref{RotCoor} we have
\begin{multline}
C_{lm}^{m^{\prime}}=\int Y^{(\xi)*}_{lm}(\theta^{\prime},\varphi^{\prime})
Y_{lm}(\theta(\theta^{\prime},\varphi^{\prime}),\varphi(\theta^{\prime},\varphi^{\prime}))
\sin \theta^{\prime} \td\theta^{\prime} \td\varphi^{\prime}=\\
\bk{Y^{(\xi)}_{lm^{\prime}}}{RY_{lm}}=
e^{im^{\prime}\gamma}\bk{Y^{(\xi)}_{lm^{\prime}}}{e^{\frac{i}{\hbar}J_{\overline y}\beta}Y_{lm}}e^{im\alpha}=:
e^{im^{\prime}\gamma}d_{mm^{\prime}}^{l}(\beta)e^{im\alpha}
\end{multline}
where $d_{mm^{\prime}}^l(\beta)$ is the Wigner $d$-function.
For our purposes $\gamma$ can always be chosen equal to $0$, so
\be
Y_{lm}(\theta(\theta^{\prime},\varphi^{\prime}),\varphi(\theta^{\prime},\varphi^{\prime}))
=e^{im\alpha}\sum_{m^{\prime}}d_{mm^{\prime}}^{l}(\beta)Y_{lm^{\prime}}^{(\xi)}(\theta^{\prime},\varphi^{\prime})
\ee
Like \cite{Podolskiy2004} we use the phase convention 
$Y_{lm}^{*}(\theta,\varphi)=(-1)^mY_{l-m}(\theta,\varphi)$
and define real-valued spherical harmonics
\be
\label{spherdef}
\begin{split}
\overline Y_{l0}&:=Y_{l0}\\ m>0:\quad\overline
Y_{lm}&:=\sqrt{2}(-1)^m\text{Re}Y_{l|m|}\\m<0:\quad\overline Y_{lm}&:=\sqrt{2}(-1)^m\text{Im}Y_{l|m|}
\end{split}\ee
Relations \gref{spherdef} can be summarized as
\be
\overline Y_{lm}=\delta_{m0}Y_{l0}+(1-\delta_{m0})\sqrt{2}(-1)^m\big[\tau(m)\text{Re}Y_{l|m|}+\tau(-m)\text{Im}Y_{l|m|}\big]
\ee

\(\tau(m)\) is the discrete Heaviside function
\be
\tau(m)=\left\{\begin{array}{ll}
1&\textrm{if \(m\ge0\)}\\
0&\textrm{if \(m<0\)}
\end{array}\right.
\ee
In the rotated coordinate system we consider ``fundamental'' matrix elements  
of a spherically symmetric operator H (this
includes H=I for simple overlaps between wavefunctions)
\be
\label{matel}
\int\overline
Y_{l_1m_1}^{(\xi)}(\theta_{1},\varphi_{1})f_1(|\vec{r}\:|)\text{H}f_2(|\vec{r}-R\vec{\xi}\:|)\overline Y_{l_2m_2}^{(\xi)}(\theta_{2},\varphi_{2})\td^3r=(l_1l_2|m_1|)\delta_{m_1m_2}
\ee
where $(l_1l_2|m_1|)$ is defined by \gref{matel}. These quantities satisfy
\be
\label{notin}
(l_1l_2|m_1|)=(-1)^{l_1-l_2}(l_2l_1|m_1|)
\ee
Here the exchange $l_{1}\leftrightarrow l_{2}$ refers to an interchange of atoms and implies
$\vec{\xi} \rightarrow -\vec{\xi}$.
Because of the relation \gref{notin} not all the matrix elements  are
independent. It is therefore sufficient to specify the
elements for which $l_1\leq l_2$; the number of independent parameters in a tight binding computer program
is thus reduced.\\
The Slater-Koster tables express a general matrix element between \emph{unrotated} functions
\be
\label{unrmatel}
\int\overline
Y_{l_1m_1}f_1\text{H}f_2\overline Y_{l_2m_2}\td^3r=\bra{l_1m_1}\text{H}\ket{l_2m_2}
\ee
in terms of the fundamental matrix elements \gref{matel} and coefficients depending on 
the angular momentum quantum numbers of the states involved. The general form of the relation
between \gref{matel} and \gref{unrmatel} according to  
\cite{Podolskiy2004} is
\be
\label{MEresult}
\begin{split}
\bra{l_1m_1}\text{H}\ket{l_2m_2}&(\alpha,\beta,R)=\sum_{m^{\prime}=1}^{l_<}
\biggl[S_{m_1m^{\prime}}^{l_1}(\alpha,\beta)S_{m_2m^{\prime}}^{l_2}(\alpha,\beta)
+T_{m_1m^{\prime}}^{l_1}(\alpha,\beta)T_{m_2m^{\prime}}^{l_2}(\alpha,\beta)\biggr]\\
&\qquad\cdot(l_1l_2|m^{\prime}|)(R)+
2A_{m_1}(\alpha)A_{m_2}(\alpha)d_{|m_1|0}^{l_1}(\beta)d_{|m_2|0}^{l_2}(\beta)(l_1l_20)(R)
\end{split}\ee
where
\be
A_m(\alpha):=\left\{ \begin{array}{ll}
(-1)^m\bigl[\tau(m)\cos(|m|\alpha)+\tau(-m)\sin(|m|\alpha)\bigr] & \textrm{if \(m\neq0\)}\\
\frac{1}{\sqrt{2}} & \textrm{if } m=0
\end{array}\right.
\ee
\be
B_m(\alpha):=\left\{\begin{array}{ll}(-1)^m\bigl[
  \tau(-m)\cos(|m|\alpha)-\tau(m)\sin(|m|\alpha)\bigr]&\textrm{if \(m\neq0\)}\\
0 & \textrm{if } m=0
\end{array}\right.\ee
\be
S_{mm^{\prime}}^l:=A_m\big[(-1)^{m^{\prime}} d_{|m|m^{\prime}}^l+d_{|m|-m^{\prime}}^l\big]
\ee
\be
T_{mm^{\prime}}^l:=B_m\big(1-\delta_{m0}\big)\big[(-1)^{m^{\prime}}d_{|m|m^{\prime}}^l
-d_{|m|-m^{\prime}}^l\big]
\ee
An explicit form of the Wigner $d$-function is
\be
\label{dwigner}
\begin{split}
d_{mm^{\prime}}^l(\beta)&=2^{-l}(-1)^{l-m^{\prime}}
\bigl[(l+m)!(l-m)!(l+m^{\prime})!(l-m^{\prime})!\bigr]^{\frac{1}{2}}\\
&\qquad\cdot\sum_{k=k_>}^{k_<}\frac{(-1)^k(1-\cos\beta)^{l-k-\frac{m+m^{\prime}}{2}}
(1+\cos\beta)^{k+\frac{m+m^{\prime}}{2}}}{k!(l-m-k)!(l-m^{\prime}-k)!(m+m^{\prime}+k)!}
\end{split}
\ee
with
$l_<=\min (l_1,l_2),\; k_<=\min (l-m,l-m^{\prime}),\; k_>=\max (0,-m-m^{\prime})$.
Expression \gref{dwigner} has been obtained from \cite{Varshal}, replacing $(\cos (\beta/2))^{2}$ with
$(1+\cos\beta)/2$ and $(\sin (\beta/2))^{2}$ with $(1-\cos\beta)/2$.
The summation limits in \gref{dwigner} are obtained from imposing the
existence condition on the factorials. They differ from \cite{Podolskiy2004} but they
assure the minimum number of terms in the sum. As is shown in \cite{voglcomment}
these conditions guarantee that the function is always defined.\\ 
We recover the classical notation of fundamental matrix elements replacing
the quantum numbers $l$ and $m$ with their spectroscopic notation
($l=\{0,1,2,\ldots\}\rightarrow\{s,p,d,\ldots\}$,
$|m|=\{0,1,2,\ldots\}\rightarrow$ $\{\sigma,\pi,\delta,\ldots\}$).
For the evaluation of $\cos(m\alpha)$ or $\sin(m\alpha)$ in $A_m$ and $B_m$
the following relations (Moivre) are useful:
\be
\begin{split}
\cos(m\alpha)=&\sum\limits_{k=0}^{\left[\frac{m}{2}\right]}(-1)^{k}\binom{m}{2k}(\sin \alpha)^{2k}
(\cos \alpha)^{m-2k}\\
\sin(m\alpha)=&\sum\limits_{k=0}^{\left[\frac{m-1}{2}\right]}(-1)^{k}\binom{m}{2k+1}(\sin \alpha)^{2k+1}
(\cos \alpha)^{m-2k-1}
\end{split}
\ee
Here $[x]$ denotes the largest integer not larger than $x$.
%
%
%
%
%
%
\section{First derivative of matrix elements}
\label{Fder}
The matrix elements have been expressed in \gref{MEresult} 
as functions of $\alpha$, $\beta$, $R$. 
In tight binding molecular dynamics simulations the derivatives of the matrix elements with 
respect to the Cartesian components of the connecting vector are required. We express
$\alpha$, $\beta$, $R$ as functions of $R^x$, $R^y$, $R^z$. As
$\beta=\arccos (R^{z}/{R})$ we have
\be
\label{Fderbeta}
\p{\beta}{R^a}=-\frac{1}{\sqrt{1-(R^z/R)^2}}\cdot
\biggl\{\frac{\delta^{za}}{R}-\frac{R^zR^a}{R^3}\biggr\}=
-\frac{\delta^{za}-(R^zR^a)/(R^2)}{\sqrt{R^{2}-(R^z)^{2}}}
\ee
and, if $R^x\neq0$, $\tan\alpha=R^{y}/R^{x}$ yields 
$\alpha=\arctan (R^{y}/R^{x})+\varphi$, where
\begin{equation*}
\varphi=\left\{\begin{array}{ll}
0& {R^x>0,R^y\geq 0}\\
{2\pi}&{R^x>0,R^y<0}\\
{\pi}&{R^x<0}
\end{array}\right.\end{equation*}

\be
\label{Fderalpha}
\p{\alpha}{R^a}=\frac{1}{1+(R^y/R^x)^{2}}\cdot\biggl\{\frac{\delta^{ya}}{R^x}
-\frac{R^y\delta^{xa}}{(R^x)^2}\biggr\}=\frac{R^x\delta^{ya}-R^y\delta^{xa}}{(R^x)^2+(R^y)^2}=
\frac{R^x\delta^{ya}-R^y\delta^{xa}}{R^{2}-(R^{z})^{2}}
\ee
The case $R^x=0$, $R^y\neq0$ implies $\cos\alpha=0$, meaning  $\alpha=\pi/2$
or $\alpha=3\pi/2$; the last two expressions in \gref{Fderalpha} 
are valid in this case, too.\\
If $R^x=0$ and $R^y=0$ the connecting vector is aligned along the $z$-axis, and $R^z=\pm R$.
Consequently $\alpha$ is undefined and there are also potential problems in \gref{Fderbeta}. 
The situation at $R^z=\pm R$ is considered in section \ref{poles}.
The expressions in the present section are valid everywhere except at the poles $R^z=\pm R$.\\
Let us introduce the notation 
$F(\alpha,\beta,R):=\bra{l_1m_1}H\ket{l_2m_2}(\alpha,\beta,R)$.
The derivative then is
\begin{multline}
\p{}{R^a}\bra{l_1m_1}\text{H}\ket{l_2m_2}=\p{F}{\alpha}\p{\alpha}{R^a}
+\p{F}{\beta}\p{\beta}{R^a}+\p{F}{R}\p{R}{R^a}=\\
\frac{R^x\delta^{ya}-R^y\delta^{xa}}{(R^x)^2+(R^y)^2}
\p{F}{\alpha}-\frac{\delta^{za}-(R^{z}R^{a})/(R^2)}{\sqrt{R^2-(R^z)^2}}\p{F}{\beta}+\frac{R^a}{R}\p{F}{R}
\end{multline}
From \gref{MEresult} the derivatives of $F$ are
\be
\label{DerivOfF}
\begin{split}
\p{F}{\alpha}&=\sum_{m^{\prime}=1}^{l_<}\biggl[\p{S_{m_1m^{\prime}}^{l_1}}{\alpha}S_{m_2m^{\prime}}^{l_2}
+S_{m_1m^{\prime}}^{l_1}\p{S_{m_2m^{\prime}}^{l_2}}{\alpha}+\p{T_{m_1m^{\prime}}^{l_1}}{\alpha}
T_{m_2m^{\prime}}^{l_2}+T_{m_1m^{\prime}}^{l_1}\p{T_{m_2m^{\prime}}^{l_2}}{\alpha}\biggr](l_1l_2|m^{\prime}|)\\&+2\de{A_{m_1}}{\alpha}A_{m_2}d_{|m_1|0}^{l_1}d_{|m_2|0}^{l_2}(l_1l_20)+2A_{m_1}\de{A_{m_2}}{\alpha}d_{|m_1|0}^{l_1}d_{|m_2|0}^{l_2}(l_1l_20)\\=&\sum_{m^{\prime}=1}^{l_<}\biggl[\p{S_{m_1m^{\prime}}^{l_1}}{\alpha}S_{m_2m^{\prime}}^{l_2}+S_{m_1m^{\prime}}^{l_1}\p{S_{m_2m^{\prime}}^{l_2}}{\alpha}+\p{T_{m_1m^{\prime}}^{l_1}}{\alpha}
T_{m_2m^{\prime}}^{l_2}+T_{m_1m^{\prime}}^{l_1}\p{T_{m_2m^{\prime}}^{l_2}}{\alpha}\biggr](l_1l_2|m^{\prime}|)\\&+2|m_1|B_{m_1}A_{m_2}d_{|m_1|0}^{l_1}d_{|m_2|0}^{l_2}(l_1l_20)+2|m_2|A_{m_1}B_{m_2}d_{|m_1|0}^{l_1}d_{|m_2|0}^{l_2}(l_1l_20)
\end{split}\ee
\be\begin{split}
\p{F}{\beta}&=\sum_{m^{\prime}=1}^{l_<}\biggl[\p{S_{m_1m^{\prime}}^{l_1}}{\beta}S_{m_2m^{\prime}}^{l_2}+S_{m_1m^{\prime}}^{l_1}\p{S_{m_2m^{\prime}}^{l_2}}{\beta}+\p{T_{m_1m^{\prime}}^{l_1}}{\beta}T_{m_2m^{\prime}}^{l_2}+T_{m_1m^{\prime}}^{l_1}\p{T_{m_2m^{\prime}}^{l_2}}{\beta}\biggr](l_1l_2|m^{\prime}|)\\
&+2A_{m_1}A_{m_2}\de{d_{|m_1|0}^{l_1}}{\beta}d_{|m_2|0}^{l_2}(l_1l_20)+2A_{m_1}A_{m_2}d_{|m_1|0}^{l_1}\de{d_{|m_2|0}^{l_2}}{\beta}(l_1l_20)
\end{split}\ee
\be\begin{split}
\p{F}{R}&=\sum_{m^{\prime}=1}^{l_<}\biggl[S_{m_1m^{\prime}}^{l_1}S_{m_2m^{\prime}}^{l_2}+T_{m_1m^{\prime}}^{l_1}T_{m_2m^{\prime}}^{l_2}\biggr]\de{(l_1l_2|m^{\prime}|)}{R}+2A_{m_1}A_{m_2}d_{|m_1|0}^{l_1}d_{|m_2|0}^{l_2}\de{(l_1l_20)}{R}
\end{split}\ee
The derivatives of $S_{mm^{\prime}}^l$ and $T_{mm^{\prime}}^l$ with respect to
$\alpha,\beta$ are given by

\be\begin{split}
\p{S_{mm^{\prime}}^l}{\alpha}=\de{A_m}{\alpha}\biggl[(-1)^{m^{\prime}}
  d_{|m|m^{\prime}}^l+d_{|m|-m^{\prime}}^l\biggr]=|m|B_m\biggl[(-1)^{m^{\prime}}
  d_{|m|m^{\prime}}^l+d_{|m|-m^{\prime}}^l\biggr]
\end{split}\ee

\be
\p{S_{mm^{\prime}}^l}{\beta}=A_m\biggl[(-1)^{m^{\prime}}\de{d_{|m|m^{\prime}}^l}{\beta}+\de{d_{|m|-m^{\prime}}^l}{\beta}\biggr]
\ee

\be\begin{split}
\p{T_{mm^{\prime}}^l}{\alpha}&=\de{B_m}{\alpha}\biggl(1-\delta_{m0}\biggr)\biggl[(-1)^{m^{\prime}}d_{|m|m^{\prime}}^l-d_{|m|-m^{\prime}}^l\biggr]\\=
&-|m|A_m\biggl(1-\delta_{m0}\biggr)\biggl[(-1)^{m^{\prime}}d_{|m|m^{\prime}}^l-d_{|m|-m^{\prime}}^l\biggr]
\end{split}\ee

\be\begin{split}
\p{T_{mm^{\prime}}^l}{\beta}=B_m\biggl(1-\delta_{m0}\biggr)\biggl[(-1)^{m^{\prime}}\de{d_{|m|m^{\prime}}^l}{\beta}-\de{d_{|m|-m^{\prime}}^l}{\beta}\biggr]
\end{split}\ee
where use has been made of
\be
\label{am}
\de{A_m}{\alpha}=|m|B_m,
\qquad
\de{B_m}{\alpha}=-|m|A_m
\ee
The derivatives of the Wigner $d$-function can be expressed as
\be
\label{beta}
\de{d_{mm^{\prime}}^l}{\beta}=\frac{1}{2}\biggl[(l+m^{\prime})(l-m^{\prime}+1)\biggr]^{\frac{1}{2}}d_{mm^{\prime}-1}^l-\frac{1}{2}\biggl[(l-m^{\prime})(l+m^{\prime}+1)\biggr]^{\frac{1}{2}}d_{mm^{\prime}+1}^l\\
\ee
and are defined for $|m^{\prime}|\le l$. 
The radial part of our function is given by the fundamental matrix elements 
defined in \gref{matel}, which are model dependent quantities. Their derivatives with respect to $R$ 
are thus also model dependent.
\section{Second derivative of matrix elements}
\label{Sder}
Relations \gref{am} and \gref{beta} are recursive relations for
computing the derivatives of $A_m$, $B_m$, $d_{mm^{\prime}}^{l}$, so in principle
we can compute any higher derivative of the matrix elements. 
The tight binding approaches we are aware of require derivatives with respect to 
the nuclear positions up to second order. Below we list the corresponding expressions for 
these second order derivatives, valid at all points except the poles, which will 
be discussed in section \ref{poles}. 

\begin{multline}
\label{GenSecDer}
\ppp{}{R^b}{R^a}\bra{l_1m_1}\text{H}\ket{l_2m_2}=\ppp{\alpha}{R^b}{R^a}
\p{F}{\alpha}+\p{\alpha}{R^a}\biggl(\p{\alpha}{R^b}\pp{F}{\alpha}+\p{\beta}{R^b}\ppp{F}{\beta}{\alpha}
+\frac{R^b}{R}\ppp{F}{R}{\alpha}\biggr)+\\
\ppp{\beta}{R^b}{R^a}\p{F}{\beta}+\p{\beta}{R^a}\biggl(\p{\alpha}{R^b}\ppp{F}{\alpha}{\beta}
+\p{\beta}{R^b}\pp{F}{\beta}+\frac{R^b}{R}\ppp{F}{R}{\beta}\biggr)+\\
\biggl(\frac{\delta^{ab}}{R}-\frac{R^aR^b}{R^3}\biggr)\p{F}{R}
+\frac{R^a}{R}\biggl(\p{\alpha}{R^b}\ppp{F}{\alpha}{R}+\p{\beta}{R^b}\ppp{F}{\beta}{R}
+\frac{R^b}{R}\pp{F}{R}\biggr)
\end{multline}
where the derivatives involved are given by
\be
\ppp{\beta}{R^b}{R^a}=\frac{\delta^{zb}R^aR^2+
\delta^{ab}R^zR^2-2R^zR^aR^b}{R^4\sqrt{R^2-(R^z)^2}}
-\frac{\bigl(R^zR^a-R^2\delta^{za}\bigr)
\bigl(R^b-R^z\delta^{zb}\bigr)}{\bigl[R^2-(R^z)^2\bigr]^{\frac{3}{2}}R^2}
\ee
\be
\ppp{\alpha}{R^b}{R^a}=\frac{\delta^{xb}\delta^{ya}-\delta^{yb}\delta^{xa}}{(R^x)^2+(R^y)^2}
-2\frac{\bigl(R^x\delta^{ya}-R^y\delta^{xa}\bigr)\bigl(R^x\delta^{xb}
+R^y\delta^{yb}\bigr)}{\bigl[(R^x)^2+(R^y)^2\bigr]^{2}}
\ee

\be
\pp{F}{R}=2A_{m_1}A_{m_2}d_{|m_1|0}^{l_1}d_{|m_2|0}^{l_2}\dde{}{R}(l_1l_20)
+\sum_{m^{\prime}=1}^{l_<}\biggl[S_{m_1m^{\prime}}^{l_1}S_{m_2m^{\prime}}^{l_2}
+T_{m_1m^{\prime}}^{l_1}T_{m_2m^{\prime}}^{l_2}\biggr]\dde{}{R}(l_1l_2|m^{\prime}|)
\ee

\begin{multline}
\pp{F}{\alpha}=2\biggl[2|m_1m_2|B_{m_1}B_{m_2}-(m_1^2+m_2^2)A_{m_1}A_{m_2}\biggr]
d_{|m_1|0}^{l_1}d_{|m_2|0}^{l_2}(l_1l_20)\\
+\sum_{m^{\prime}=1}^{l_<}\biggl[\biggl(\pp{}{\alpha}S_{m_1m^{\prime}}^{l_1}\biggr)S_{m_2m^{\prime}}^{l_2}
+S_{m_1m^{\prime}}^{l_1}\biggl(\pp{}{\alpha}S_{m_2m^{\prime}}^{l_2}\biggr)
+2\biggl(\p{}{\alpha}S_{m_1m^{\prime}}^{l_1}\biggr)\biggl(\p{}{\alpha}S_{m_2m^{\prime}}^{l_2}\biggr)\\
+\biggl(\pp{}{\alpha}T_{m_1m^{\prime}}^{l_1}\biggr)T_{m_2m^{\prime}}^{l_2}
+T_{m_1m^{\prime}}^{l_1}\biggl(\pp{}{\alpha}T_{m_2m^{\prime}}^{l_2}\biggr)
+2\biggl(\p{}{\alpha}T_{m_1m^{\prime}}^{l_1}\biggr)\biggl(\p{}{\alpha}T_{m_2m^{\prime}}^{l_2}\biggr)
\biggr](l_1l_2|m^{\prime}|)
\end{multline}

\be\begin{split}
\pp{F}{\beta}=&2A_{m_1}A_{m_2}\biggl[\biggl(\dde{}{\beta}d_{|m_1|0}^{l_1}\biggr)d_{|m_2|0}^{l_2}+2\biggl(\de{}{\beta}d_{|m_1|0}^{l_1}\biggr)\biggl(\de{}{\beta}d_{|m_2|0}^{l_2}\biggr)+d_{|m_1|0}^{l_1}\biggl(\dde{}{\beta}d_{|m_2|0}^{l_2}\biggr)\biggr](l_1l_20)\\&+\sum_{m^{\prime}=1}^{l_<}\biggl[\biggl(\pp{}{\beta}S_{m_1m^{\prime}}^{l_1}\biggr)S_{m_2m^{\prime}}^{l_2}+S_{m_1m^{\prime}}^{l_1}\biggl(\pp{}{\beta}S_{m_2m^{\prime}}^{l_2}\biggr)+2\biggl(\p{}{\beta}S_{m_1m^{\prime}}^{l_1}\biggr)\biggl(\p{}{\beta}S_{m_2m^{\prime}}^{l_2}\biggr)\\&+\biggl(\pp{}{\beta}T_{m_1m^{\prime}}^{l_1}\biggr)T_{m_2m^{\prime}}^{l_2}+T_{m_1m^{\prime}}^{l_1}\biggl(\pp{}{\beta}T_{m_2m^{\prime}}^{l_2}\biggr)+2\biggl(\p{}{\beta}T_{m_1m^{\prime}}^{l_1}\biggr)\biggl(\p{}{\beta}T_{m_2m^{\prime}}^{l_2}\biggr)\biggr](l_1l_2|m^{\prime}|)
\end{split}\ee

\be\begin{split}
\ppp{F}{\beta}{\alpha}=&2\biggl(|m_1|B_{m_1}A_{m_2}+|m_2|A_{m_1}B_{m_2}\biggr)\biggl[\biggl(\de{}{\beta}d_{|m_1|0}^{l_1}\biggr)d_{|m_2|0}^{l_2}+d_{|m_1|0}^{l_1}\biggl(\de{}{\beta}d_{|m_2|0}^{l_2}\biggr)\biggr](l_1l_20)\\&+\sum_{m^{\prime}=1}^{l_<}\biggl[\biggl(\ppp{}{\alpha}{\beta}S_{m_1m^{\prime}}^{l_1}\biggr)S_{m_2m^{\prime}}^{l_2}+S_{m_1m^{\prime}}^{l_1}\biggl(\ppp{}{\alpha}{\beta}S_{m_2m^{\prime}}^{l_2}\biggr)+\biggl(\p{}{\beta}S_{m_1m^{\prime}}^{l_1}\biggr)\biggl(\p{}{\alpha}S_{m_2m^{\prime}}^{l_2}\biggr)\\&+\biggl(\p{}{\alpha}S_{m_1m^{\prime}}^{l_1}\biggr)\biggl(\p{}{\beta}S_{m_2m^{\prime}}^{l_2}\biggr)+\biggl(\ppp{}{\alpha}{\beta}T_{m_1m^{\prime}}^{l_1}\biggr)T_{m_2m^{\prime}}^{l_2}+T_{m_1m^{\prime}}^{l_1}\biggl(\ppp{}{\alpha}{\beta}T_{m_2m^{\prime}}^{l_2}\biggr)\\&+\biggl(\p{}{\beta}T_{m_1m^{\prime}}^{l_1}\biggr)\biggl(\p{}{\alpha}T_{m_2m^{\prime}}^{l_2}\biggr)+\biggl(\p{}{\alpha}T_{m_1m^{\prime}}^{l_1}\biggr)\biggl(\p{}{\beta}T_{m_2m^{\prime}}^{l_2}\biggr)\biggr](l_1l_2|m^{\prime}|)
\end{split}\ee

\be\begin{split}
\ppp{F}{R}{\alpha}=&2\biggl(|m_1|B_{m_1}A_{m_2}+|m_2|A_{m_1}B_{m_2}\biggr)d_{|m_1|0}^{l_1}d_{|m_2|0}^{l_2}\de{}{R}(l_1l_20)+\sum_{m^{\prime}=1}^{l_<}\biggl[\biggl(\p{}{\alpha}S_{m_1m^{\prime}}^{l_1}\biggr)S_{m_2m^{\prime}}^{l_2}\\&+S_{m_1m^{\prime}}^{l_1}\biggl(\p{}{\alpha}S_{m_2m^{\prime}}^{l_2}\biggr)+\biggl(\p{}{\alpha}T_{m_1m^{\prime}}^{l_1}\biggr)T_{m_2m^{\prime}}^{l_2}+T_{m_1m^{\prime}}^{l_1}\biggl(\p{}{\alpha}T_{m_2m^{\prime}}^{l_2}\biggr)\biggr]\de{}{R}(l_1l_2|m^{\prime}|)
\end{split}\ee

\be\begin{split}
\ppp{F}{R}{\beta}=&2A_{m_1}A_{m_2}\biggl[\biggl(\de{}{\beta}d_{|m_1|0}^{l_1}\biggr)d_{|m_2|0}^{l_2}+d_{|m_1|0}^{l_1}\biggl(\de{}{\beta}d_{|m_2|0}^{l_2}\biggr)\biggr]\de{}{R}(l_1l_20)\\&+\sum_{m^{\prime}=1}^{l_<}\biggl[\biggl(\p{}{\beta}S_{m_1m^{\prime}}^{l_1}\biggr)S_{m_2m^{\prime}}^{l_2}+S_{m_1m^{\prime}}^{l_1}\biggl(\p{}{\beta}S_{m_2m^{\prime}}^{l_2}\biggr)+\biggl(\p{}{\beta}T_{m_1m^{\prime}}^{l_1}\biggr)T_{m_2m^{\prime}}^{l_2}\\&+T_{m_1m^{\prime}}^{l_1}\biggl(\p{}{\beta}T_{m_2m^{\prime}}^{l_2}\biggr)\biggr]\de{}{R}(l_1l_2|m^{\prime}|)
\end{split}\ee

\be
\pp{S_{mm^{\prime}}^l}{\alpha}=\dde{A_m}{\alpha}\biggl[(-1)^{m^{\prime}}
  d_{|m|m^{\prime}}^l+d_{|m|-m^{\prime}}^l\biggr]=-m^2A_m\biggl[(-1)^{m^{\prime}}
  d_{|m|m^{\prime}}^l+d_{|m|-m^{\prime}}^l\biggr]
\ee

\be
\pp{S_{mm^{\prime}}^l}{\beta}=A_m\biggl[(-1)^{m^{\prime}}\dde{d_{|m|m^{\prime}}^l}{\beta}+\dde{d_{|m|-m^{\prime}}^l}{\beta}\biggr]
\ee

\be\begin{split}
\pp{T_{mm^{\prime}}^l}{\alpha}=&\dde{B_m}{\alpha}\biggl(1-\delta_{m0}\biggr)\biggl[(-1)^{m^{\prime}}d_{|m|m^{\prime}}^l-d_{|m|-m^{\prime}}^l\biggr]\\=&-m^2B_m\biggl(1-\delta_{m0}\biggr)\biggl[(-1)^{m^{\prime}}d_{|m|m^{\prime}}^l-d_{|m|-m^{\prime}}^l\biggr]
\end{split}\ee

\be\begin{split}
\pp{T_{mm^{\prime}}^l}{\beta}=B_m\biggl(1-\delta_{m0}\biggr)\biggl[(-1)^{m^{\prime}}\dde{d_{|m|m^{\prime}}^l}{\beta}-\dde{d_{|m|-m^{\prime}}^l}{\beta}\biggr]
\end{split}\ee

\be\begin{split}
\ppp{S_{mm^{\prime}}^l}{\beta}{\alpha}=\ppp{S_{mm^{\prime}}^l}{\alpha}{\beta}=&\de{A_m}{\alpha}\biggl[(-1)^{m^{\prime}}\de{d_{|m|m^{\prime}}^l}{\beta}+\de{d_{|m|-m^{\prime}}^l}{\beta}\biggr]\\=&|m|B_m\biggl[(-1)^{m^{\prime}}\de{d_{|m|m^{\prime}}^l}{\beta}+\de{d_{|m|-m^{\prime}}^l}{\beta}\biggr]
\end{split}\ee

\be\begin{split}
\ppp{T_{mm^{\prime}}^l}{\beta}{\alpha}=\ppp{T_{mm^{\prime}}^l}{\alpha}{\beta}=
&\de{B_m}{\alpha}\biggl(1-\delta_{m0}\biggr)
\biggl[(-1)^{m^{\prime}}\de{d_{|m|m^{\prime}}^l}{\beta}
-\de{d_{|m|-m^{\prime}}^l}{\beta}\biggr]\\
=&-|m|A_m\biggl(1-\delta_{m0}\biggr)\biggl[(-1)^{m^{\prime}}\de{d_{|m|m^{\prime}}^l}{\beta}
-\de{d_{|m|-m^{\prime}}^l}{\beta}\biggr]
\end{split}\ee
\section{The poles $R=\pm R^{z}$}
\label{poles}
\begin{figure}
\begin{center}
\includegraphics[width=7cm]{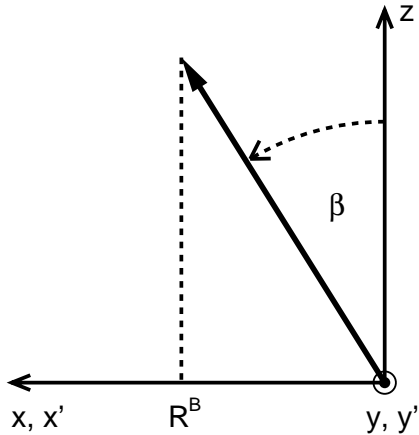}
\end{center}
\caption{Geometry for the derivatives at the poles. For $\partial/\partial R^{x}$ the $x$- and $z$-
axis are as shown, the $y$-axis, as is also indicated, is pointing towards the reader. 
For $\partial/\partial R^{y}$ a rotation about the $z$-axis by $\pi/2$ is necessary, and instead of the
$x$- and $y$-axis we obtain the $x^{\prime}$- and $y^{\prime}$-axis, the latter pointing towards the 
reader. As far as $\beta$ is concerned, both situations are identical. The reader should note that
$R^{B}$ is always nonnegative by construction.}
\label{restricted}
\end{figure}

If the second atom is located on the $z$-axis, the azimuthal angle $\alpha$ cannot be defined. The 
derivative of $\alpha$ with respect to $R^{x}$ or $R^{y}$ diverges as the connecting vector approaches the 
$z$-axis, i.e if $|R^{z}| \rightarrow R$. In the same limit $\partial\beta/\partial R^{z} \rightarrow 0$ 
and the derivatives of $\beta$ with respect to $R^{x}$ or $R^{y}$ remain bounded, but do not converge.
Related behaviour is found for the second derivatives as well.
These problems are of course due to a well known {\it coordinate} singularity, but, given that we need 
some expressions at the poles (to type into a computer, eventually), we have to look at this case. 
A further issue to be addressed is how to handle the dependence of $F$ and of its derivatives on $\alpha$ 
at the poles, i.e. what value of $\alpha$ to plug into the expressions where this angle is not defined.
\newline
In order to solve both problems it is helpful to recall what exactly the Euler angles mean, see section 
\ref{ME}. If both atoms are on the $z$-axis, a rotation about this axis does not change the matrix element
between any two orbitals, because both orbitals get rotated by the same angle. Therefore there is no 
$\alpha$-dependence for the matrix element, as can also be verified by evaluating \gref{MEresult} in the limit
$\beta \rightarrow 0,\pi$. A partial derivative is the limit of a differential quotient. At the poles 
this means 
the difference between the value of the matrix element at some point $P$ and the value at the pole, 
divided by the difference in the respective Cartesian coordinate we are considering, taken in the limit that
$P$ approaches the pole. But, because our point of reference is a pole, the choice of $P$, even if at an
``infinitesimal'' distance from the pole, determines the angle $\alpha$ by which to rotate around the
$z$-axis, {\it before} a rotation around the thus obtained $y^{\prime}$-axis by the angle $\beta$.
Therefore, regardless along which direction in the $xy$-plane we move away from the pole, we are always in
the situation shown in figure \ref{restricted}; this is also evident from the azimuthal symmetry about the
pole. 
For $\beta$ we thus have in this case $\beta=\arccos (R^{z}/R)$ with $R=\sqrt{(R^{z})^{2}+(R^{B})^{2}}$, 
and it follows
\be
\label{betalimit}
\p{\beta}{R^{x}}=\p{\beta}{R^{y}}=\p{\beta}{R^{B}}=\frac{1}{\sqrt{R^{2}-(R^{z})^{2}}}
\frac{R^{z}R^{B}}{R^{2}}=\frac{R^{z}}{R^{2}}\rightarrow \pm \frac{1}{R}
\ee
where the limit is for $\beta \rightarrow 0,\pi$, respectively.
The result
\be
\p{\beta}{R^{z}}=-\frac{1}{R}\sqrt{1-\left(\frac{R^{z}}{R}\right)^{2}}\rightarrow 0
\ee
is as expected.
Note that the total derivative of $\beta$ exists everywhere, but is not continuous at the poles; thus there
will be no total second derivative of $\beta$ at the poles.
\newline
The meaning of the Euler angles leads to the following choice for the values of $\alpha$ to be used in 
partial first derivatives of $F$ at the poles:
$F$ and $\partial F / \partial R$ do not depend on $\alpha$ at the poles, as follows from \gref{MEresult}, 
so its choice is arbitrary.
For expressions involving $\partial F / \partial \beta$, the choice to be made is
$\alpha=0$ for $x$-derivatives and $\alpha=\pi/2$ for $y$-derivatives. The same choices of $\alpha$,
applied to \gref{Fderbeta} for $\partial \beta / \partial R^{x}$ and $\partial \beta / \partial R^{y}$, 
respectively, reproduce the result \gref{betalimit}.   
\newline
Any deviation from the pole can be entirely expressed in terms of changes in $R$ and $\beta$, $\alpha$ 
only enters to fix the direction of the deviation. Partial derivatives with respect to $\alpha$, i.e. 
changing $\alpha$ while keeping $\beta$ and $R$ fixed, can't be made sense of at the poles. They do not occur. 
As furthermore we have $\partial R/\partial R^{a}=\pm \delta^{za}$ at $R^{z}=\pm R$, the general 
expression for the first derivatives of a matrix element at $\beta=0,\pi$ respectively, is
\be
\label{PoleFirst}
\p{}{R^{a}}\bra{l_{1},m_{1}} H \ket{l_{2},m_{2}}=\pm \delta^{za}\p{F}{R}
\pm \delta^{xa}\frac{1}{R}\p{F}{\beta}|_{\alpha=0}
\pm \delta^{ya}\frac{1}{R}\p{F}{\beta}|_{\alpha=\pi/2}
\ee
Turning to the second partial derivatives, we find that all but one can be expressed in terms of
derivatives of $F$, $\beta$ and $R$. The problematic one is the mixed $R^{x}R^{y}$-derivative, 
because it involves two different directions in the $xy$-plane and thus two conflicting values 
of $\alpha$ to be used in $F$ and its derivatives. All other second partial derivatives involve at 
most one of the variables $R^{x}$ or $R^{y}$, and the reasoning for the first partial derivatives, as
illustrated in figure \ref{restricted}, can be carried over.
We find at the poles
\begin{displaymath}
\pp{\beta}{(R^{x})}=0,\quad \pp{\beta}{(R^{y})}=0,\quad \pp{\beta}{(R^{z})}=0,
\end{displaymath}
\begin{displaymath}
\ppp{\beta}{R^{x}}{R^{z}}=\ppp{\beta}{R^{y}}{R^{z}}=\ppp{\beta}{R^{z}}{R^{x}}=\ppp{\beta}{R^{z}}{R^{y}}=
-\frac{1}{R^{2}}
\end{displaymath}
As in the case of the first derivatives, derivatives with respect to $\alpha$ cannot be made sense 
of at the poles. 
We have
\be
\begin{split}
\pp{}{(R^{z})}\bra{l_{1},m_{1}} H \ket{l_{2},m_{2}}&=\pp{F}{R}\\
\ppp{}{R^{z}}{R^{x}}\bra{l_{1},m_{1}} H \ket{l_{2},m_{2}}&=-\frac{1}{R^{2}}\p{F}{\beta}|_{\alpha=0}+
\frac{1}{R}\ppp{F}{\beta}{R}|_{\alpha=0}\\
\ppp{}{R^{z}}{R^{y}}\bra{l_{1},m_{1}} H \ket{l_{2},m_{2}}&=-\frac{1}{R^{2}}\p{F}{\beta}|_{\alpha=\pi/2}+
\frac{1}{R}\ppp{F}{\beta}{R}|_{\alpha=\pi/2}
\end{split}
\ee
where the mixed derivatives are symmetric. 

For the mixed second derivative with respect to $R^{x}$ and $R^{y}$
we have to think of the Cartesian form of the expression \gref{MEresult}. There, from each coefficient
multiplying a fundamental matrix element $(l_{1}l_{2}|m|)$ we extract, if possible, a factor
$\sin\alpha\cos\alpha(\sin\beta)^{2}$; this corresponds to $(R^{x}R^{y})/(R^{2})$. 
It is easily seen that at the poles only those terms contribute to the derivative in question where this
extraction is possible and where the rest of the coefficient after the extraction does not vanish.
If we do this
analysis, we find that the only nonvanishing derivatives are at $\beta=0$:
\begin{multline}
\ppp{}{R^{x}}{R^{y}}\bra{l_{1},0}H\ket{l_{2},-2}=
\frac{1}{2\sqrt{2}R^{2}}\sqrt{(l_{2}+2)(l_{2}+1)l_{2}(l_{2}-1)}\;(l_{1}l_{2}0)
-\\
(1-\delta_{l_{1}0})\frac{1}{\sqrt{2}R^{2}}
\sqrt{l_{1}(l_{1}+1)(l_{2}+2)(l_{2}-1)}\;
(l_{1}l_{2}1)+\\
(1-\delta_{l_{1}0})(1-\delta_{l_{1}1})\frac{1}{2\sqrt{2}R^{2}}
\sqrt{(l_{1}+2)(l_{1}+1)l_{1}(l_{1}-1)}\;(l_{1}l_{2}2)
\end{multline}
and
\begin{multline}
\ppp{}{R^{x}}{R^{y}}\bra{l_{1},+1}H\ket{l_{2},-1}=\\\frac{1}{R^{2}}\left\{
\frac{1}{2}\sqrt{l_{1}(l_{1}+1)l_{2}(l_{2}+1)}\;(l_{1}l_{2}0)-\frac{1}{4}
[l_{1}(l_{1}+1)+l_{2}(l_{2}+1)]\;(l_{1}l_{2}1)\right\}
\end{multline}
At $\beta=\pi$ we obtain $(-1)^{l_{1}+l_{2}}\times$ these values.
\newline

We think that the expressions we have given in the previous sections and completed in this one are 
suitable for implementation in a TB-based molecular dynamics code. Though the expressions are somewhat
lengthy, they are well structured and can be coded by successive function calls. The functions $A_{m}$, 
$B_{m}$, and $d^{l}_{mm^{\prime}}$ are expressed explicitly in the code, $S^{l}_{mm^{\prime}}$, 
$T^{l}_{mm^{\prime}}$ and their derivatives with respect to $\alpha$ are implemented in terms of 
these quantities. Derivatives with respect to $\beta$ are evaluated calling derivatives of the
Wigner $d$-function, which themselves are stated as combinations of Wigner $d$-functions.
\section*{Acknowledgements}
This work was supported by EPSRC under grant GR/S80165.

\end{document}